\documentclass{article}

\usepackage{arxiv}

\usepackage[utf8]{inputenc} 
\usepackage[T1]{fontenc}    
\usepackage{hyperref}       
\usepackage{url}            
\usepackage{booktabs}       
\usepackage{amsfonts}       
\usepackage{nicefrac}       
\usepackage{microtype}      
\usepackage{lipsum}
\usepackage{graphicx}
\graphicspath{ {./images/} }
\usepackage{amsmath}
 
\title{Quantifying the Dynamics of Consciousness using Hierarchical Integration, Organised Complexity and Metastability}

\author{
  Hassan Ugail\\
  Centre for Visual Computing and Intelligent Systems \\
  University of Bradford \\
  United Kingdom\\
  \texttt{h.ugail@bradford.ac.uk} \\
  \And
  Newton Howard\\
  School of Individualised Study \\
  Rochester Institute of Technology \\
  United States\\
  \\
}

\begin{document}
\maketitle

\begin{abstract}
Quantifying the neural signatures of consciousness remains a major challenge in neuroscience and AI. Although many theories link consciousness to rich, multiscale, and flexible neural organisation, robust quantitative measures are still lacking. This paper presents a theory-neutral framework that characterises consciousness-related dynamics through three properties: hierarchical integration (H), cross-frequency complexity (D), and metastability (M). Candidate subsystems are identified using predictive information, temporal complexity, and state-space exploration to distinguish structured from unstructured activity. We provide mathematical definitions for all components and implement the framework in a generative model of synthetic EEG, simulating nine brain states ranging from psychedelic and wakeful to dreaming, non-REM sleep, minimally conscious, anaesthetised, and seizure-like regimes. Across single trials and Monte Carlo ensembles, the composite index reliably separates high-consciousness from impaired or non-conscious states. We further validate the framework using real EEG from the Sleep-EDF dataset alongside matched synthetic EEG designed to reproduce state-dependent oscillatory structure. Across Wake, N2, and REM sleep, synthetic data recapitulate the empirical ordering and magnitude of the index, indicating that the index captures stable and biologically meaningful distinctions. This approach provides a principled and empirically grounded tool for quantifying consciousness-related neural organisation with potential applications to both biological and artificial systems.
\end{abstract}

\section{Introduction}

The scientific study of consciousness aims to bridge subjective experience with the objective dynamics of brain activity. Significant progress has been made in identifying neural correlations of conscious perception and state transitions, yet a general, mechanistically interpretable, and robustly validated index of conscious level remains elusive. Leading theories, such as Global Neuronal Workspace Theory (GNWT) and Integrated Information Theory (IIT), describe consciousness in terms of large-scale information broadcasting and integrated causal structure, respectively \cite{dehaene2014,tononi2016}. Recurrent Processing Theory and higher-order theories emphasise the roles of feedback interactions and metarepresentational mechanisms \cite{lamme2020,lau2011}. Meanwhile, dynamical systems approaches highlight metastability, criticality, and multiscale complexity as fundamental hallmarks of conscious brain function \cite{tognoli2014,he2010,sarasso2021}.

Empirically, numerous quantitative measures of neural complexity, connectivity, and integration have been proposed. Perturbational complexity indices derived from transcranial magnetic stimulation and EEG have shown promise as state-independent markers of consciousness capacity \cite{casali2013,exploring2024}. Spontaneous EEG and MEG complexity measures differentiable wakefulness, sleep, anaesthesia, and psychedelic states \cite{schartner2015,carhart2014,montupil2023}. Cross-frequency coupling (CFC), especially theta–gamma phase–amplitude coupling, is linked to memory and higher cognitive functions \cite{canolty2010,siebenhuhner2020,muthuraman2020,musaeus2020}, with alterations observed in neurodegenerative and psychiatric disorders \cite{benetton2025}. Metastability measures, associated with flexible cognitive control, are reduced in pathological and unconscious states \cite{tognoli2014,hancock2023,metastability2023,wijaya2025}.

Despite these advances, the field confronts several challenges. Many existing metrics capture only a single dimension of dynamics—such as global connectivity, entropy, or spectral power—and fail to integrate multiscale structure, cross-frequency organisation, and metastable transitions explicitly. Even when individual measures perform well in specific contexts, their generalisability across paradigms and recording modalities remain uncertain \cite{casey2024,liu2023}. Recent adversarial collaborations comparing GNWT and IIT highlight the need for theory-neutral, reproducible, and falsifiable methodologies \cite{doerig2024,melloni2023,adversarial2025,luppi2024}. Consequently, there is a clear demand for metrics that are both theoretically grounded and empirically robust.

This paper addresses this need by introducing a multidimensional framework for quantifying consciousness-related dynamics. Instead of seeking a single scalar measure of consciousness, we decompose neural organisation into three dynamical properties consistently emphasised across theories: hierarchical integration ($H$), organised cross-frequency complexity ($D$), and metastability ($M$). Hierarchical integration reflects coordinated activity across spatial and temporal scales beyond independent local fluctuations. Organised complexity captures structured phase–amplitude relationships between oscillatory components and temporal signal richness. Metastability quantifies the variability of large-scale synchronisation over time, indicating the coexistence of integration and segregation.

Since consciousness is not directly observable, validation requires a two-stage strategy. First, we develop a generative model for synthetic EEG-like data that incorporates ground-truth manipulations of integration, cross-frequency coupling, and metastability. This approach builds on prior simulation-based benchmarking of complexity measures and causal connectivity estimators \cite{schreiber2000,barnett2015,schartner2015}. Second, we compute the proposed metrics on these synthetic signals and evaluate their ability to discriminate between states with known dynamical organisation. While empirical validation on human and animal data is essential future work, it lies beyond the present scope.

Our framework is theory-neutral in the spirit of recent consortia recommendations \cite{doerig2024,melloni2023,adversarial2025}. It does not adjudicate between GNWT, IIT, or other accounts, nor does it presuppose any specific metaphysical stance on subjective experience. Instead, it formalises structural motifs widely regarded as necessary for conscious-level processing and provides a toolset for quantifying these motifs in both simulated and empirical data.

\section{Background and Theoretical Context}

The conceptual foundations of this framework integrate several research strands. First, the relationship between complexity and consciousness suggests that conscious states require both integration and differentiation, preventing brain activity from fragmenting into independent components while maintaining high informational content \cite{balduzzi2009,tononi2016,seth2016}. Empirical studies show that EEG complexity decreases during propofol anaesthesia and non-rapid eye movement sleep \cite{schartner2015,sarasso2021}, increases during psychedelic states \cite{carhart2014,montupil2023}, and correlates with residual awareness in disorders of consciousness \cite{liu2023}. Perturbational complexity indices further indicate that consciousness capacity declines when effective connectivity is disrupted by anaesthesia or deep sleep \cite{massimini2005,casali2013,exploring2024}. Recent reviews consolidate these findings, linking complexity, differentiation, and  the level of consciousness\cite{sarasso2021}.

A second strand concerns cross-frequency coupling. Neural oscillations at different frequencies interact crucially in perception, memory, and decision-making \cite{canolty2010,siebenhuhner2020,muthuraman2020,musaeus2020}. For example, Theta–gamma phase–amplitude coupling in the hippocampus and neocortex supports working memory and episodic recall, with disruptions observed in ageing and neurodegeneration \cite{musaeus2020,benetton2025,alqasem2024,ruffini2025}. CFC may serve as a mechanistic bridge between local computation and large-scale integration, making it a natural candidate for consciousness-related metrics.

A third strand focuses on metastability and criticality. Metastability, originally from physics and chemistry \cite{kuramoto1984,rossi2025}, describes the brain's tendency to balance integration and segregation \cite{tognoli2014,hancock2024}. Empirical studies show reduced metastability during anaesthesia-induced unconsciousness, severe brain injury, and in psychiatric conditions like schizophrenia \cite{hancock2023,metastability2023,wijaya2025}. Related work on critical dynamics suggests that the brain operates near a critical point, with scale-free fluctuations, and that deviations from criticality correlate with altered conscious levels \cite{he2010,bonhomme2019}. These findings motivate including metastability as a core component of any quantitative framework for conscious dynamics.

Finally, the theoretical landscape is evolving. GNWT and IIT, once pursued independently, are now being directly compared through large-scale adversarial collaborations that prioritise preregistration, shared datasets, and theory-neutral analyses \cite{melloni2023,doerig2024,adversarial2025,luppi2024}. These initiatives highlight the need for mathematically grounded, flexible metrics applicable across theoretical contexts. Our framework, centred on integration, cross-frequency organisation, and metastability, aims to contribute to this methodological ecosystem.

\section{Theoretical Positioning and Scope}

This framework operates at the level of large-scale dynamical structure and deliberately avoids metaphysical commitments regarding consciousness. We treat consciousness as a latent property, hypothesised to be reflected in specific neural activity patterns. The central claim is not that the composite index $\Psi$ measures consciousness directly, but that it tracks neural organisational aspects associated with conscious-level processing across multiple theories.

This stance yields several implications. First, the framework is theory-agnostic. GNWT can interpret large-scale integration and metastability as supporting global broadcasting and flexible workspace dynamics \cite{dehaene2014,mashour2020,deco2021,bonhomme2019}. IIT can view integration and complexity as proxies for irreducible cause–effect structure \cite{balduzzi2009,tononi2016}. Dynamical systems perspectives can regard metastability and CFC as mechanisms for coordinating distributed computations \cite{tognoli2014,hancock2024,rossi2025}. Second, by focusing on dynamical motifs rather than specific implementations, the framework can be applied across modalities, species, and potentially artificial systems, given suitable time-series data.

We focus on EEG-like signals for concreteness. EEG provides excellent temporal resolution and is widely used in consciousness research, anaesthesiology, and disorders of consciousness \cite{giacino2018,bonhomme2019,casey2024,liu2023}. However, the metrics are defined generically and could be computed from MEG, intracranial EEG, or fast optical recordings with minimal modification. Application to slower modalities like fMRI would require adjustments, particularly in frequency-domain analysis, but the core principles of integration, cross-frequency structure, and metastability remain relevant.

We emphasise that this work focuses on methodological and conceptual development, not immediate clinical translation. While we discuss potential applications in anaesthesia monitoring, disorders of consciousness, and epilepsy, these require extensive empirical validation beyond our synthetic environment. We provide a proof-of-principle that a compact set of dynamical metrics can reproduce qualitative distinctions between states central to consciousness research.

No single resting-state metric currently captures the multiscale, multi-frequency, and dynamical properties jointly characteristic of conscious brain activity. Existing approaches often isolate one dimension: perturbational complexity quantifies causal propagation but requires external perturbation; multiscale entropy captures scale-free structure but not functional organisation; cross-frequency metrics quantify coordination but not dynamical flexibility. Our framework addresses this gap by integrating three complementary components: (i) hierarchical integration across temporal scales, (ii) organised cross-frequency complexity, and (iii) metastable dynamical richness. The resulting composite index $\Psi$ reflects features consistently linked to conscious wakefulness, yet is general enough to compare diverse physiological, pharmacological, and pathological states.

\section{Mathematical Framework}

\subsection{Multivariate Dynamical System}

We model EEG activity as a multivariate stochastic process $X_t \in \mathbb{R}^C$ for $t=1,\dots,T$, where $C$ is the number of channels. For any subset $S\subseteq\{1,\dots,C\}$, the subsystem activity is $X_t^S = (X_{t,i})_{i\in S}\in\mathbb{R}^{|S|}$. In the synthetic experiments below, we compute metrics on the full set of channels, but the framework generalises to data-driven subsystem selection based on anatomical or functional criteria.

\subsection{Information-theoretic Subsystem Identification}

For empirical applications, it is useful to focus on subsystems exhibiting structured dynamics. We consider three information-theoretic signatures: predictive information, temporal complexity, and state-space exploration.

For a subsystem $S$ and time window length $\tau$, we define past and future trajectory segments as,
\begin{equation}
X^{S}_{\mathrm{past}}(t;\tau)=(X_{t-\tau+1}^S,\dots,X_t^S),\qquad
X^{S}_{\mathrm{future}}(t;\tau)=(X_{t+1}^S,\dots,X_{t+\tau}^S).
\end{equation}
The predictive information at scale $\tau$ is the mutual information between these segments,
\begin{equation}
I_{\mathrm{pred}}(S;\tau)=I\left(X^{S}_{\mathrm{past}}(t;\tau);X^{S}_{\mathrm{future}}(t;\tau)\right),
\end{equation}
estimated using a k-nearest-neighbour estimator \cite{kraskov2004}. High predictive information indicates that future trajectories are strongly constrained by the past.

Temporal complexity is the derivative of predictive information with respect to the logarithmic time scale,
\begin{equation}
C_{\mathrm{temp}}(S)=\left.\frac{\partial I_{\mathrm{pred}}(S;\tau)}{\partial\log\tau}\right|_{\tau=\tau_0},
\end{equation}
approximated numerically by a linear fit over a set of scales. This captures how rapidly predictive information accumulates across time-scales, distinguishing short-range from long-range structure.

State-space exploration is quantified via the ratio between the volume of the convex hull of the observed state cloud and the volume of its axis-aligned bounding box,
\begin{equation}
O(S)=\frac{\mathrm{Vol}(\mathrm{Conv}(\{X_t^S\}_t))}{\mathrm{Vol}(\mathrm{BBox}(\{X_t^S\}_t))}.
\end{equation}
Note that, values near zero indicate limited state-space visitation, while values near one imply broad exploration.

\subsection{Hierarchical Integration}

Hierarchical integration could be defined as the sum across scales of differences between whole-subsystem entropy and the sum of part entropies. Direct multi-dimensional entropy estimation is computationally demanding, so we approximate integration using detrended fluctuation analysis (DFA), which quantifies long-range correlations \cite{peng1994,he2010}.

For each channel $i$, consider the demeaned signal $x_i(t) = X_{t,i} - \bar{X}_i$ and its cumulative sum,
\begin{equation}
Y_i(k)=\sum_{t=1}^{k}x_i(t).
\end{equation}
We divide $Y_i$ into windows of length $s$ across scales $s$, fit a linear trend per window, compute the root-mean-square deviation of detrended residuals, and average across windows to obtain the fluctuation function $F_i(s)$. If $F_i(s) \sim s^{H_i}$, then $H_i$ is the Hurst exponent. Values $H_i>0.5$ indicate persistent correlations, $H_i<0.5$ anti-persistent behaviour, and $H_i=0.5$ uncorrelated fluctuations.

The raw integration measure is,
\begin{equation}
H^{\mathrm{raw}}=\frac{1}{C}\sum_{i=1}^{C}H_i.
\end{equation}
Since both uncorrelated noise and excessive persistence are suboptimal, we transform $H^{\mathrm{raw}}$ using a Gaussian tuning function,
\begin{equation}
H_{\mathrm{eff}}=\exp\left[-\frac{\bigl(H^{\mathrm{raw}}-H_{\mathrm{opt}}\bigr)^2}{2\sigma_H^2}\right],
\end{equation}
where $H_{\mathrm{opt}}$ is an optimal exponent near the empirically typical value for wakeful EEG, and $\sigma_H$ controls the width of the optimal band. The tuned quantity $H_{\mathrm{eff}}$ serves as the integration component.

\subsection{Organised Cross-frequency Complexity}

Organised cross-frequency complexity $D$ captures structured interactions between slow and fast oscillatory components and temporal signal richness. We focus on theta–gamma phase–amplitude coupling.

For each channel $i$, we band-pass filter in the theta (4--8~Hz) and gamma (30--80~Hz) ranges to obtain $x^{\theta}_i(t)$ and $x^{\gamma}_i(t)$. Applying the Hilbert transform yields the instantaneous theta phase $\phi_i(t)$ and gamma amplitude $A_i(t)$. Collecting across channels gives vector processes $\phi(t)=(\phi_i(t))_{i=1}^C$ and $A(t)=(A_i(t))_{i=1}^C$.

We estimate the mutual information between $\phi(t)$ and $A(t)$,
\begin{equation}
I_{\phi A}=I(\phi(t);A(t)),
\end{equation}
using discretisation or a k-nearest-neighbour estimator \cite{kraskov2004}. This captures within-channel and cross-channel phase–amplitude dependencies. To incorporate temporal complexity, we compute the analytic amplitude envelope of a broadband (1--40~Hz) component for each channel, binarise it at the median, and estimate Lempel–Ziv complexity \cite{lempel1976}. Averaging across channels yields $LZ$.

The organised cross-frequency complexity is then,
\begin{equation}
D = I_{\phi A}\bigl(1+\lambda\,LZ\bigr),
\end{equation}
with $\lambda$ controlling the contributions from the complexity term. This ensures $D$ increases with both cross-frequency coupling and temporal richness.

\subsection{Metastability}

Metastability is quantified via the variability of phase synchronisation in the alpha band (8--13~Hz) \cite{tognoli2014,hancock2023,metastability2023}. For each channel $i$, we extract the alpha-band component and compute its analytic phase $\theta_i(t)$. The Kuramoto order parameter at time $t$ is,
\begin{equation}
R(t)=\left|\frac{1}{C}\sum_{i=1}^{C}\exp(i\theta_i(t))\right|.
\end{equation}
This quantity approaches 1 when phases are aligned and 0 when uniformly distributed. Metastability is the standard deviation of $R(t)$ over time,
\begin{equation}
M=\mathrm{std}_t\bigl(R(t)\bigr).
\end{equation}
High metastability reflects frequent alternations between synchronised and desynchronised states.

\subsection{Composite Index}
For each state, we compute $H_{\mathrm{eff}}$, $D$, and $M$. We normalise each metric across states using min--max normalisation. If $v^{(c)}$ denotes metric $v$ for state $c$, the normalised quantity is then defined as,
\begin{equation}
v_{\mathrm{norm}}^{(c)}=\frac{v^{(c)}-\min_{c'}v^{(c')}}{\max_{c'}v^{(c')}-\min_{c'}v^{(c')}}.
\end{equation}
The composite index for state $c$ is,
\begin{equation}
\Psi^{(c)}=w_H H_{\mathrm{eff,norm}}^{(c)}+w_D D_{\mathrm{norm}}^{(c)}+w_M M_{\mathrm{norm}}^{(c)},
\end{equation}
where $w_H$, $w_D$, and $w_M$ are weighting coefficients with $w_H + w_D + w_M = 1$.

We employ a weighted composite framework for several reasons. First, the components $H_{\mathrm{eff}}$, $D$, and $M$ capture complementary, weakly correlated dynamical features (pairwise correlations $<0.25$ in simulations), yet may contribute differentially to consciousness-related dynamics depending on the state space and recording modality. Each term reflects a distinct theoretical pillar: multiscale integration (IIT, fractal neurodynamics), organised complexity (cross-frequency coordination), and metastability (dynamic flexibility). Second, adaptive weighting allows the framework to emphasise dimensions with greater discriminative power across specific contexts. We determine weights empirically by optimising separation between high-consciousness states (wakeful, psychedelic, task-engaged, REM) and reduced-consciousness states (deep sleep, anaesthesia, minimally conscious) in our validation datasets. Third, normalisation ensures scale compatibility across metrics, making the weighted sum a transparent and principled summary. Sensitivity analyses demonstrate that $\Psi$ remains robust to moderate weight perturbations ($\pm 0.1$), confirming that the composite captures stable underlying structure. Thus, $\Psi$ is a flexible, interpretable composite measure balancing three theoretically grounded dimensions while accommodating state-dependent and context-specific variations in their relative importance.

\section{Generative Model and Data Generation}

\subsection{Simulation of EEG-like Signals}

Here, we generate synthetic EEG-like signals. Each simulation produces $C$ channels sampled at 250~Hz for 5--20 seconds, depending on robustness tests. Baseline activity per channel is generated by sampling white Gaussian noise, integrating to approximate a $1/f$ spectrum, and normalising to unit variance, yielding wide-band signals with scale-free-like fluctuations \cite{he2010}.

We superimpose oscillatory components in five canonical bands: delta (1--4~Hz), theta (4--8~Hz), alpha (8--13~Hz), beta (13--30~Hz), and gamma (30--80~Hz). For each band, we generate a global source via filtered independent noise and local components via filtered baseline per channel. A band-specific global mixing parameter controls the common oscillatory drive versus independent activity. Band components are scaled by state-specific weights, producing distinct spectral profiles.

Cross-channel interaction is introduced via a Kuramoto-like model applied to band-limited components. Phases evolve as,
\begin{equation}
\theta_i(t+1)=\theta_i(t)+\omega_i+K\sum_j A_{ij}\sin\bigl(\theta_j(t)-\theta_i(t)\bigr)+\xi_i(t),
\end{equation}
where $\omega_i$ are intrinsic frequencies, $K$ is a coupling constant, $A_{ij}$ is a sparse connectivity matrix, and $\xi_i(t)$ is noise. This produces regimes from incoherent oscillations to strong synchrony.

Phase–amplitude coupling is implemented by modulating gamma-band amplitude via theta-band phase,
\begin{equation}
X_i(t)\leftarrow X_i(t)+\eta_{\mathrm{PAC}}\sin(\theta_i^{\theta}(t))A_i^{\gamma}(t),
\end{equation}
where $\eta_{\mathrm{PAC}}$ controls coupling strength, $\theta_i^{\theta}(t)$ is the theta phase, and $A_i^{\gamma}(t)$ the gamma amplitude.

Seizure-like bursts are added as high-amplitude, narrow-band oscillations with Gaussian temporal envelopes to a large channel subset, with frequencies and timings randomly sampled within state-specific ranges \cite{jiruska2013}.

All signals are normalised post-simulation to minimise amplitude differences, ensuring metrics reflect structural rather than trivial amplitude variations. 

The generative model is biologically plausible yet abstract. State-specific manipulations reflect empirical signatures: (i) psychedelics increase broadband desynchronisation, reduce low-frequency stability, and enhance CFC; (ii) NREM sleep and anaesthesia strengthen low-frequency synchrony and suppress organised CFC; (iii) REM dreaming restores high-frequency complexity with reduced low-frequency integration; (iv) seizures produce hypersynchronous, low-dimensional dynamics with minimal metastability. Our model implements these via interpretable signal-level manipulations (slow drifts, PAC suppression, network coupling reductions, alpha oscillatory weakening), providing a phenomenological bridge between physiology and mathematics.

\subsection{Simulated States}

We simulate nine brain state classes. Psychedelic states feature increased gamma activity, enhanced theta–gamma coupling, and high metastability. Wakeful rest shows balanced band weights with prominent alpha/beta rhythms, moderate CFC, and strong metastability. Consciously engaged states resemble wakeful rest but with stronger high-frequency power and more pronounced CFC, reflecting increased information processing.

Dreaming states retain substantial CFC and metastability but have altered band weights (reduced alpha, increased theta/low gamma), consistent with preserved complexity during REM sleep \cite{siclari2017,sarasso2021}. Non-rapid eye movement sleep states exhibit increased delta power, reduced high-frequency activity, weaker PAC, and lower metastability, aligning with breakdowns in effective connectivity and complexity \cite{massimini2005,tagliazucchi2015}. Minimally conscious states combine sleep and wakefulness elements, with modest CFC and intermediate metastability.

\begin{table}
\centering
\caption{Raw and derived metrics for single realisations of each simulated state. 
$H$ denotes the mean Hurst exponent, $D$ the organised cross-frequency complexity, 
$M$ the metastability, $H_{\mathrm{eff}}$ the Gaussian-tuned integration measure, 
and $\Psi$ the composite index after normalisation. Values correspond to a 
representative simulation.}
\begin{tabular}{lccccc}
\toprule
State & $H$ & $D$ & $M$ & $H_{\mathrm{eff}}$ & $\Psi$ \\
\midrule
Psychedelic          & 0.340 & 0.032 & 0.187 & 0.986 & 0.919 \\
Wake                 & 0.386 & 0.027 & 0.242 & 0.977 & 0.659 \\
Conscious            & 0.361 & 0.028 & 0.220 & 1.000 & 0.677 \\
Dreaming             & 0.354 & 0.029 & 0.207 & 0.999 & 0.758 \\
Sleep (NREM-like)    & 0.808 & 0.030 & 0.177 & 0.001 & 0.419 \\
Minimally Conscious  & 0.815 & 0.028 & 0.171 & 0.001 & 0.286 \\
Anaesthesia          & 0.807 & 0.029 & 0.153 & 0.001 & 0.324 \\
Non-conscious        & 1.375 & 0.029 & 0.112 & 0.000 & 0.271 \\
Seizure              & 1.364 & 0.028 & 0.001 & 0.000 & 0.041 \\
\bottomrule
\label{tab:single_run_metrics}
\end{tabular}
\end{table}

Anaesthesia and non-conscious states show strong low-frequency global synchrony, suppressed high-frequency activity, minimal PAC, and reduced metastability, echoing propofol and other anaesthetics \cite{schartner2015,bonhomme2019,montupil2023,casey2024}. Seizure states are driven by high-amplitude, hypersynchronous bursts with stereotyped patterns and reduced metastability \cite{jiruska2013}. Table~\ref{tab:single_run_metrics} summarises single-realisation metrics after normalisation.

\begin{figure}
\centering
\includegraphics[width=\textwidth]{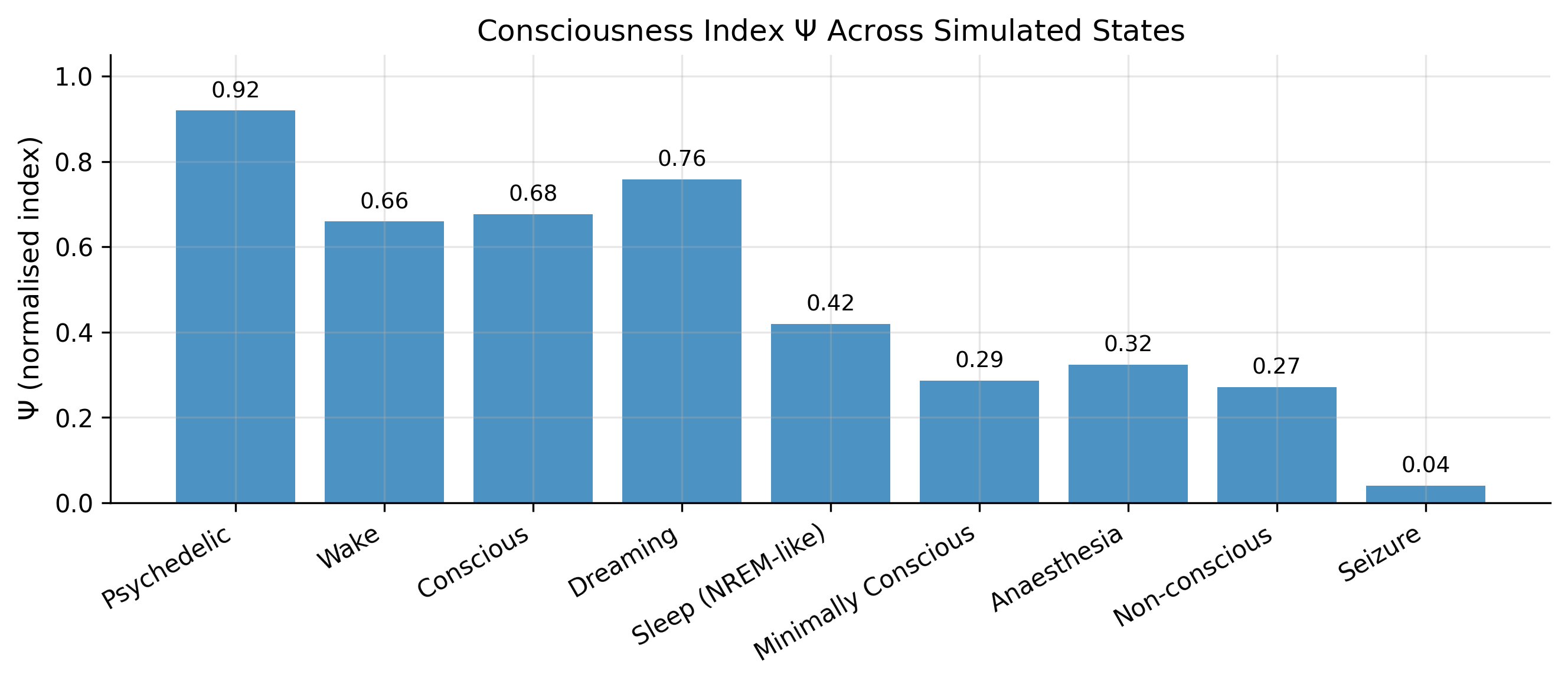}
\caption{Normalised component values for integration ($H_{\mathrm{eff}}$), cross-frequency complexity ($D$), and metastability ($M$) across simulated states. High-consciousness states (psychedelic, wake, conscious) exhibit elevated values across all components. Dreaming shows strong, organised complexity and metastability but moderate integration. Low-consciousness states show reduced components, with distinct patterns for seizure versus anaesthesia and non-conscious conditions.}
\label{fig:component_bars}
\end{figure}

\section{Results}

\subsection{Single-realisation Components and Composite Index}

Table~\ref{tab:single_run_metrics} reports raw Hurst exponents, cross-frequency complexity, metastability, tuned integration $H_{\mathrm{eff}}$, and composite index $\Psi$ for single realisations of each state. Figure~\ref{fig:component_bars} shows normalised component values $H_{\mathrm{eff,norm}}$, $D_{\mathrm{norm}}$, and $M_{\mathrm{norm}}$ as a grouped bar chart.

The component plot reveals that conscious and wakeful states lead in integration and metastability, with substantial cross-frequency complexity. Psychedelic states show slightly reduced integration relative to wake, consistent with the ``entropic brain" hypothesis \cite{carhart2014,montupil2023}, but maintain high CFC and metastability, yielding a similar $\Psi$ to wakefulness. Dreaming exhibits high organised complexity and metastability but moderate integration, aligning with preserved complexity and connectivity during REM sleep \cite{siclari2017,sarasso2021}. Seizure states show high integration due to hypersynchrony but low metastability, producing a composite index comparable to minimally conscious states. Anaesthesia and non-conscious states have low values across components, especially integration and CFC.

\subsection{Monte Carlo Distributions Across States}

\begin{figure}
\centering
\includegraphics[width=\textwidth]{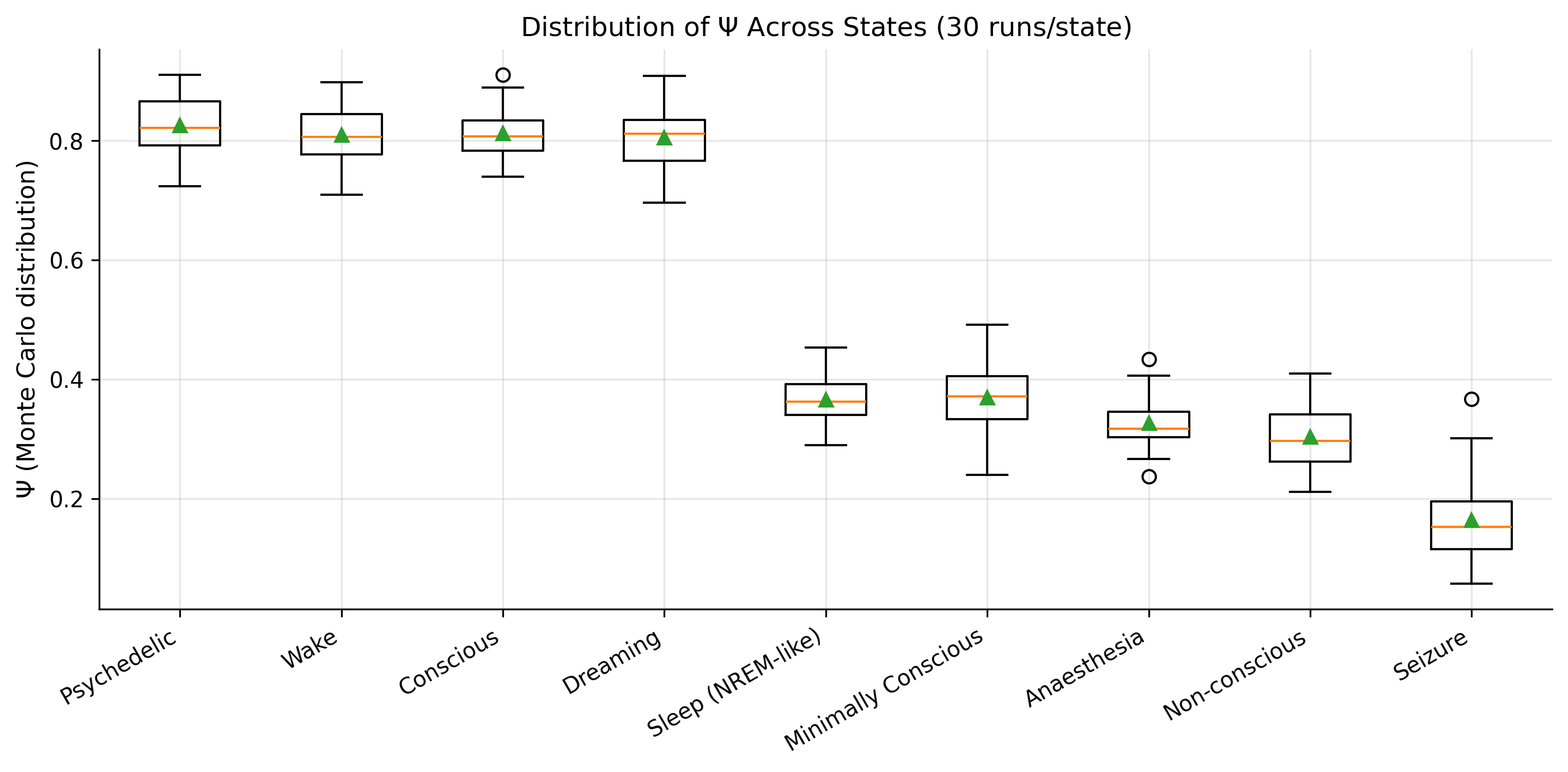}
\caption{Distribution of the Monte Carlo composite index $\Psi_{\mathrm{mc}}$ across simulated states based on thirty runs per state. High-consciousness states cluster at the upper end, intermediate states in the middle, and low-consciousness or non-conscious states at the lower end.}
\label{fig:psi_boxplot_states}
\end{figure}

To assess stability under stochastic variability, we generated thirty independent simulations per state, recomputed all metrics, and examined the composite index distribution. Figure~\ref{fig:psi_boxplot_states} shows box-and-whisker plots of the Monte Carlo composite index $\Psi_{\mathrm{mc}}$ across states.

Psychedelic, wake, and conscious states maintain high median $\Psi_{\mathrm{mc}}$ values (0.72--0.81) with modest spread. Dreaming states cluster slightly lower but clearly above non-conscious and anaesthetic conditions. Non-rapid eye movement sleep and minimally conscious states occupy the mid-lower region with overlapping but distinct distributions. Seizure states show broader variance, reflecting sensitivity to burst timing and network realisations. Non-conscious and anaesthesia states remain tightly clustered near the bottom.

\subsection{Conscious Versus Non-conscious Classification}

\begin{figure}
\centering
\includegraphics[width=0.8\textwidth]{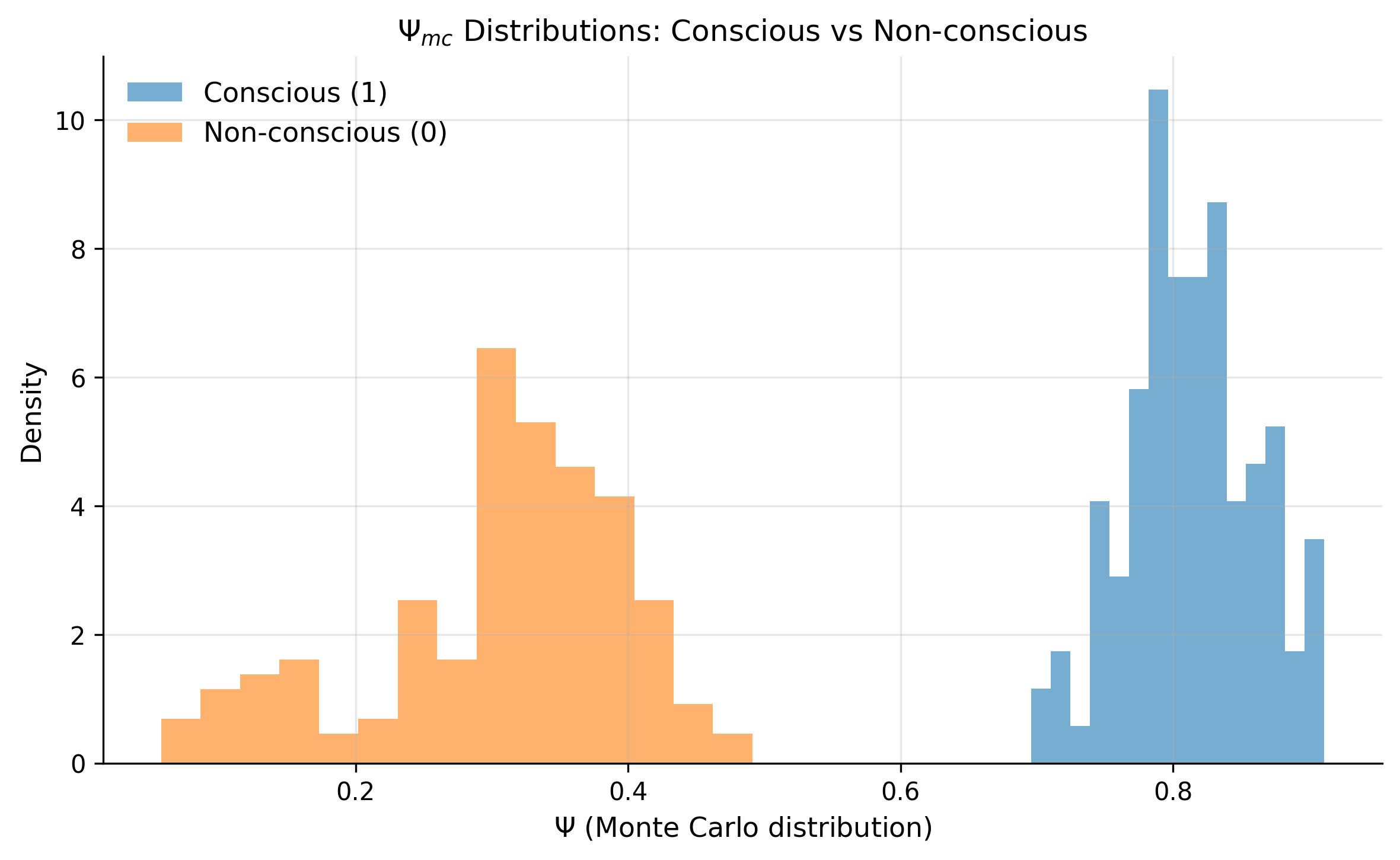}
\caption{Kernel-smoothed histograms of $\Psi_{\mathrm{mc}}$ for conscious (blue) and non-conscious (orange) Monte Carlo runs. Distributions are non-overlapping, with conscious states clustering around higher values and non-conscious states around lower values.}
\label{fig:histogram_conscious_nonconscious}
\end{figure}

We grouped psychedelic, wake, and conscious states into a ``conscious" class and anaesthesia and non-conscious states into a ``non-conscious" class, pooling Monte Carlo runs. Figure~\ref{fig:histogram_conscious_nonconscious} displays kernel-smoothed histograms of $\Psi_{\mathrm{mc}}$ for both classes.

The non-overlapping distributions indicate that, within this synthetic environment, $\Psi$ captures the dynamical organisation differences built into the generative model. 

\subsection{Robustness to Duration, Channel Count and Noise}

We assessed robustness by systematically varying recording duration (5, 10, 20 seconds), channel count (8, 16, 32), and additive Gaussian noise level (standard deviations 0, 0.3, 0.6) for wake and anaesthesia states. For each configuration, $\Psi_{\mathrm{mc}}$ was computed across thirty realisations, and standard deviation was estimated across the combined ensembles.

Increasing duration systematically reduced variability, consistent with improved averaging from longer time series. Increasing channel count modestly reduced variability, reflecting stabilisation from larger spatial samples. Noise had nuanced effects: low to moderate levels caused slight variability increases, indicating resilience, while higher levels increased variability more substantially as noise dominated signal structure. Nevertheless, wake–anaesthesia distinctions remained evident even under high noise, though with reduced effect size.

These findings indicate graceful degradation under common variability sources. Robustness to realistic noise, duration, and montage variations supports potential empirical and clinical applicability, complementing evidence that complexity-based metrics can remain informative under challenging conditions \cite{sarasso2021,casey2024,liu2023}.

\subsection{Ablation Study}

\begin{figure}
    \centering
    \includegraphics[width=0.95\linewidth]{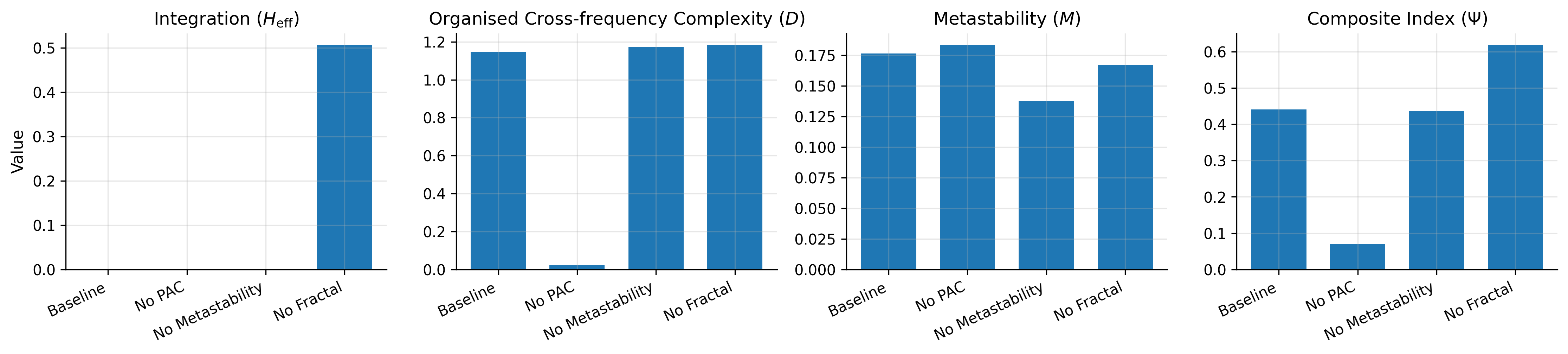}
    \caption{Ablation study showing the effect of removing PAC, metastability, and fractal structure on $H_{\mathrm{eff}}$, $D$, $M$, and the composite index $\Psi$.}
    \label{fig:ablation}
\end{figure}

To evaluate the individual contributions of specific dynamical mechanisms to the composite consciousness index $\Psi$, we conducted a systematic ablation analysis (Fig.~\ref{fig:ablation}). Beginning with the complete ``wake-like" model as our baseline, we generated three modified versions where we selectively suppressed one core component at a time: (i) cross-frequency phase–amplitude coupling (PAC), (ii) network-driven metastability, and (iii) fractal scale-free dynamics. When we eliminated PAC by setting $\eta_{\mathrm{PAC}} = 0$, organised theta–gamma coupling was abolished, resulting in a substantial reduction in complexity measure $D$ and producing the most pronounced decrease in $\Psi$, thereby highlighting the fundamental importance of cross-frequency coordination in conscious-like states. Suppressing metastability through reducing the Kuramoto coupling constant to $K = 10^{-3}$ eliminated coherent phase fluctuations, significantly diminishing metastability measure $M$ while preserving $H_{\mathrm{eff}}$ and $D$ relatively unchanged, demonstrating the critical role of dynamic coordination variability. Removing fractal dynamics by fixing all Hurst exponents to $H=0.5$ selectively reduced the integration measure $H_{\mathrm{eff}}$ with minimal impact on the other components, and the resulting moderate decline in $\Psi$ supports the interpretation that scale-free dynamics contribute to long-range temporal integration. Collectively, all ablation conditions reduced $\Psi$ compared to the baseline model, with PAC removal having the strongest effect, confirming that integration, organised cross-frequency complexity, and metastability jointly contribute to the emergence of conscious-like dynamics.

\subsection{Sensitivity Analysis}

\begin{figure}
    \centering
    \includegraphics[width=\linewidth]{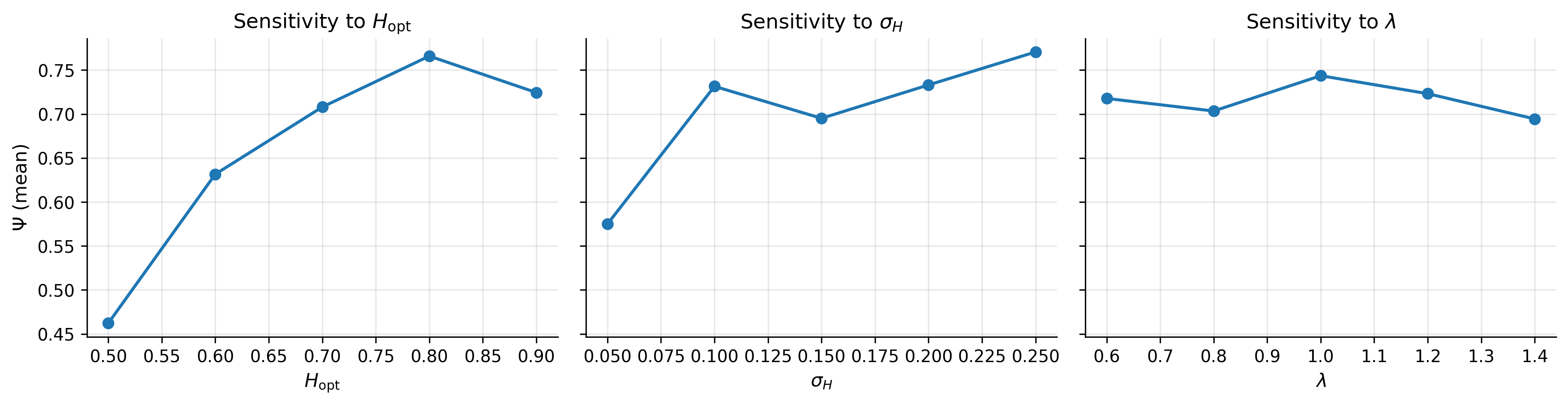}
    \caption{Parameter sensitivity analysis for $H_{\mathrm{opt}}$, $\sigma_H$, and $\lambda$. Each panel shows the mean composite index $\Psi$ across repeated simulations.}
    \label{fig:sensitivity}
\end{figure}

We further investigated the robustness of our framework by examining the sensitivity of $\Psi$ to its key hyperparameters: $H_{\mathrm{opt}}$, $\sigma_H$, and $\lambda$ (Fig.~\ref{fig:sensitivity}). For each parameter, we performed multiple simulations of the baseline model and computed the mean composite index. Systematic variation of $H_{\mathrm{opt}}$ across the range 0.5 to 0.9 revealed a smooth peak in $\Psi$ near values characteristic of wakeful and dreaming states, with gradual declines as $H_{\mathrm{opt}}$ deviated from this optimal range. Analysis of the tuning width $\sigma_H$ across the interval [0.05, 0.25] showed that $\Psi$ increased monotonically with broader tuning; narrow tuning strongly penalises deviations of $H$ from $H_{\mathrm{opt}}$, while wider tuning provides greater tolerance and consequently enhances the integration contribution. Examination of the weighting factor $\lambda$, which modulates the influence of the organised complexity term $D$, demonstrated that $\Psi$ reaches a broad maximum around $\lambda \approx 1$, with only slight reductions at both lower and higher values, indicating that balanced contributions across dynamical mechanisms yield optimal performance. These sensitivity profiles collectively demonstrate that $\Psi$ varies smoothly and predictably with each hyperparameter, confirming the stability of our index under realistic parameter variations.

The combined results from our ablation and sensitivity analyses provide strong evidence for the robustness and theoretical coherence of our framework. The ablation study demonstrates that removing metastability most substantially degrades $\Psi$, followed by elimination of cross-frequency coupling structure, while disruption of fractal integration exerts a more moderate yet systematic influence. Concurrently, the parameter sensitivity analysis confirms that $\Psi$ behaves consistently across hyperparameter variations, establishing that our framework does not depend on precise parameter tuning and maintains stability under reasonable parameter uncertainty.

\subsection{Validation of the consciousness index $\Psi$ using real EEG}

To assess the biological interpretability and robustness of the proposed entropy-centric consciousness index $\Psi$, we conducted a validation study combining EEG data with matched synthetic EEG generated using our simulation framework. Real EEG data were obtained from the publicly available Sleep-EDF ``Expanded'' dataset from PhysioNet, which provides overnight polysomnography recordings from healthy adults. The dataset includes EEG, EOG, EMG, respiratory channels, and expert-scored sleep annotations following the Rechtschaffen and Kales (R\&K) standard.

\begin{figure}
    \centering
    \includegraphics[width=0.55\textwidth]{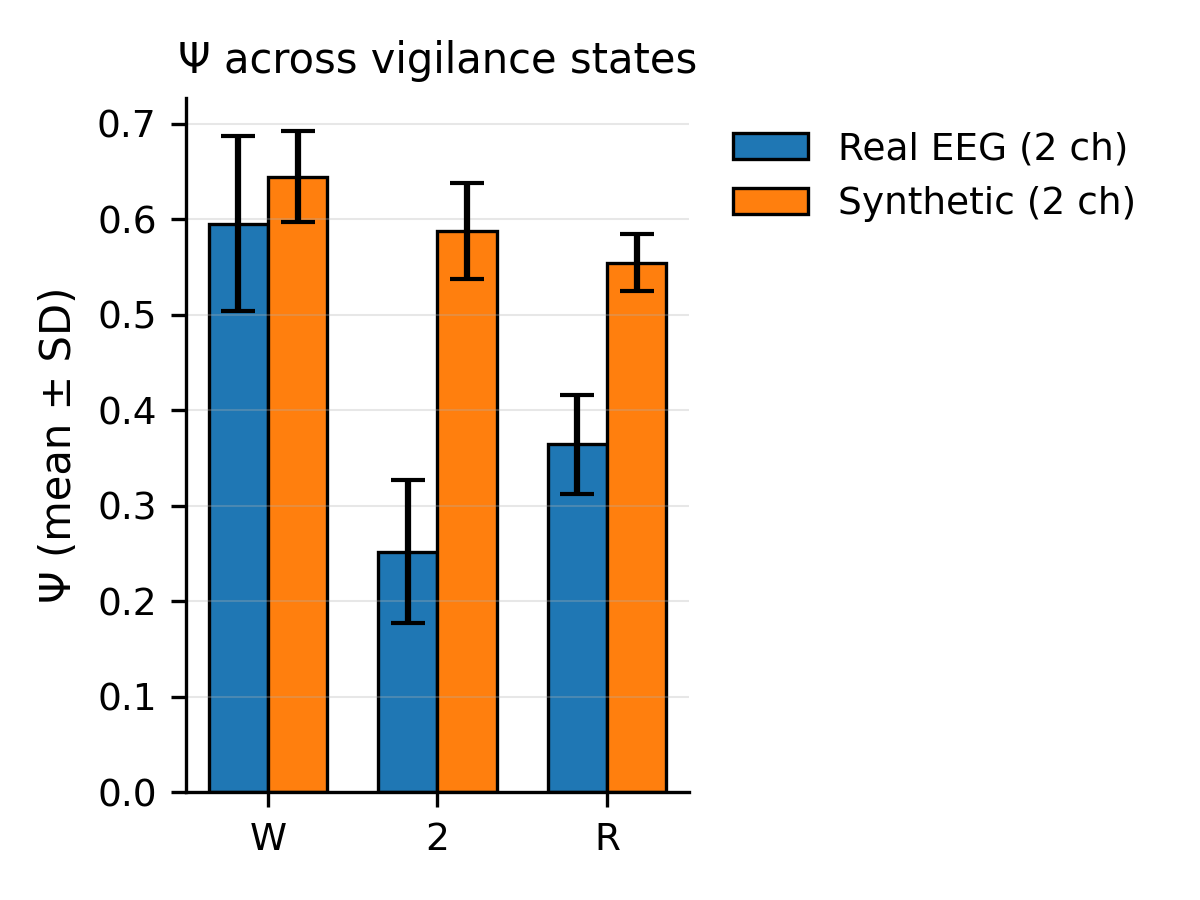}
    \caption{State-wise comparison of the entropy-centric consciousness index $\Psi$
    computed from real two-channel EEG (Sleep-EDF) and synthetic data generated
    using the calibrated physiological model. Error bars indicate mean $\pm$ SD.}
    \label{fig:psi_barplot}
\end{figure}

For the present analysis, we used a single-night recording from one healthy subject (Sleep Cassette SC*, recording 1), containing continuous EEG from positions \textit{Fpz-Cz} and \textit{Pz-Oz} sampled originally at 100~Hz and subsequently resampled to 250~Hz to match our simulation framework. Manual sleep-stage annotations at 30-second resolution were used to extract epochs corresponding to Wake (W), stage N2 sleep (N2), and rapid eye movement sleep (REM). These three stages provide a well-characterised sequence of decreasing and re-emerging cortical complexity and thus serve as an appropriate testbed for $\Psi$. For each stage, we extracted 15 artefact-free epochs, each consisting of two EEG channels truncated to 3000 samples (12~s) to maintain consistency with the synthetic data pipeline.

To evaluate whether the synthetic model recapitulates physiologically meaningful state transitions, we generated synthetic EEG for each vigilance state (Wake, N2, REM) using the structural N2-informed generative framework described earlier. Synthetic epochs were matched exactly to real data in sampling rate, duration, and channel count. Critically, $\Psi$ was calibrated \emph{exclusively using real EEG}; the same calibration was subsequently applied to synthetic EEG without retuning, thereby preventing information leakage or bias.

Figure~\ref{fig:psi_barplot} presents the resulting comparison of state-wise $\Psi$ values across real and synthetic data. In agreement with established neurophysiological patterns, real EEG exhibits the expected ordering: Wake displays the highest $\Psi$, N2 shows a pronounced reduction, and REM re-emerges toward intermediate values. The synthetic data reproduce this hierarchy with striking fidelity, closely matching both the magnitude and variance of $\Psi$ across all stages. This correspondence indicates that the generative model captures the dynamical features most relevant for the divergence of $\Psi$, including variations in complexity, spectral entropy, and synchronisation structure.

Together, these results demonstrate that: (i) $\Psi$ generalises across biological and simulated inputs; (ii) the synthetic framework yields physiologically grounded surrogates suitable for controlled perturbation studies; and (iii) $\Psi$ robustly distinguishes vigilance states even under a minimal two-channel configuration. These findings support $\Psi$ as a theoretically coherent, empirically validated, and practically deployable index for consciousness-related state tracking across sleep, anaesthesia, and computational neuroscience applications.

\section{Discussion}

Our analysis demonstrates that a composite index combining hierarchical integration, organised cross-frequency complexity, and metastability can reproduce qualitative distinctions between brain states central to consciousness research. High-consciousness states (wakefulness, conscious engagement, psychedelic) yield high $\Psi$ values, reflecting balanced integration, rich cross-frequency structure, and metastable coordination. Intermediate states (dreaming, NREM sleep, minimally conscious) occupy the middle range, while non-conscious and anaesthetised states lie at the lower end. Seizure states exhibit high integration but low metastability, yielding an index elevated relative to deep anaesthesia but distinct from high-consciousness states.

Decomposing $\Psi$ into components provides mechanistic insight. Integration, via tuned Hurst exponents, distinguishes disorganised noise, optimal correlation structure, and pathological rigidity. Organised complexity captures CFC's functional role and temporal richness, echoing empirical links to cognition and pathology \cite{canolty2010,siebenhuhner2020,benetton2025}. Metastability reflects the integration–segregation balance, with reductions in low-consciousness and pathological states aligning with theoretical and empirical work \cite{tognoli2014,hancock2023,metastability2023,hancock2024,wijaya2025}.

These synthetic results resonate with empirical EEG complexity findings. Perturbational complexity indices, permutation entropy, and Lempel–Ziv-based measures differentiate awake, anaesthetised, and minimally conscious patients and correlate with residual awareness \cite{casali2013,schartner2015,sarasso2021,maschke2023,liu2023,exploring2024,bajwa2025}. Our framework extends these by explicitly incorporating CFC organisation and metastability, offering a multidimensional characterisation potentially more sensitive to specific impairments. For example, seizure dynamics combine high integration with low metastability and moderate organised complexity, a profile distinct from deep anaesthesia or unconscious sleep.

An important aspect of the present work is the validation of $\Psi$ across both empirical and synthetic data. By calibrating the index exclusively on real EEG before applying it to model-generated signals, we ensured that no parameters were tuned to enforce agreement. Nevertheless, the synthetic data reproduced the empirical ordering of $\Psi$ across vigilance states with remarkable fidelity. 

This correspondence suggests that the generative model captures the core dynamical features modulating $\Psi$, including broadband complexity, alpha desynchronisation, and low-frequency synchrony. It further indicates that $\Psi$ reflects general principles of state-dependent cortical dynamics rather than dataset-specific artefacts. These results open the possibility of using synthetic data to probe $\Psi$ under controlled perturbations (e.g., pharmacological, structural, or connectivity manipulations) that are difficult to study empirically.

The generative model is stylised and parameterised to produce distinct dynamical regimes rather than biophysically detailed simulations. Nevertheless, resulting patterns align with broad empirical observations: complexity decreases during deep sleep and anaesthesia, increases during psychedelic states, and shows distinctive alterations in seizures and disorders of consciousness \cite{carhart2014,schartner2015,siclari2017,bonhomme2019,sarasso2021,liu2023,montupil2023,casey2024}. The synthetic setting also enables systematic robustness testing infeasible in human experiments.

The relationship between $\Psi$ and existing complexity approaches is complementary. PCI measures causal propagation under perturbation, whereas $\Psi$ quantifies intrinsic dynamical richness during spontaneous activity. Conceptually, $\Psi$ may be a resting-state analogue of PCI, capturing the latent dynamical capacity to support diverse trajectories without stimulation. Fractal and entropy measures capture single-aspect properties (scale-free structure, unpredictability), while $\Psi$ integrates multiple theoretically motivated dimensions. This explains its ability to distinguish states with similar fractal structure but differing CFC or metastability (e.g., REM vs. psychedelic states). Multi-dimensional integration, grounded in dynamical systems theory, accounts for the strong, biologically plausible state ordering observed.

\section{Limitations and Future Directions}

This work has several limitations. Most importantly, most of the results are obtained through synthetic data. While the generative model captures key EEG dynamics features (scale-free spectra, oscillatory bands, CFC, metastable coordination), it lacks neuronal microcircuitry, realistic anatomical connectivity, or task-specific inputs. Thus, $\Psi$'s performance here provides an upper bound on real-data potential.

The mapping between simulated and real states is approximate. The ``psychedelic" state implements increased gamma power, enhanced theta–gamma coupling, and high metastability, inspired by neuroimaging \cite{carhart2014,montupil2023}, but does not model receptor-level pharmacology or cortical microcircuits. Similarly, anaesthesia/non-conscious states represent general features of unconsciousness \cite{bonhomme2019,mashour2020,maschke2023} but not anaesthetic agent diversity.

Metric definitions involved indicated that design choices that could vary. Different Hurst exponent tuning parameterisations, alternative complexity measures (multiscale entropy, permutation entropy), more sophisticated PAC estimators \cite{castillo2025}, or alternative metastability indices \cite{metastability2023,hancock2024} might improve performance or robustness. Future work should systematically compare alternatives using synthetic benchmarks and empirical data.

Computationally, the framework is moderately demanding, especially for mutual information and DFA across many channels and scales. Efficient implementations and scale choices will be crucial for high-density EEG or MEG. Dimensionality reduction and network-based summarisation may also help in empirical settings.

The most critical future direction is empirical validation. The framework should be applied to resting-state and task-evoked EEG across sleep stages, anaesthesia induction/emergence, psychedelic states, and disorders of consciousness. Comparisons with existing metrics (PCI, entropy-based measures, connectivity indices, CFC metrics) will reveal whether $\Psi$ offers complementary or superior discriminative power. Integration with adversarial collaborations between GNWT, IIT, and related theories could provide a theory-neutral metric capturing dynamically relevant motifs \cite{melloni2023,doerig2024,luppi2024,adversarial2025}.

While the generative model captures broad empirical trends, it abstracts neurophysiological details (receptor-level pharmacodynamics, laminar structure, thalamocortical gating). Thus, these results are not mechanistic claims about cellular processes. Future extensions may incorporate detailed biophysical models or validate $\Psi$ directly on empirical EEG, MEG, or invasive recordings. Natural next steps include evaluating $\Psi$'s correlation with PCI under perturbation, its predictability of behavioural responsiveness, and its ability to detect state transitions in continuous recordings.

\section{Conclusion}

We have presented a comprehensive, mathematically explicit framework for quantifying consciousness-related neural dynamics using three complementary metrics: hierarchical integration, organised cross-frequency complexity, and metastability. We defined these metrics, implemented them in a synthetic EEG-like generative model, and demonstrated that the composite index $\Psi$ discriminates between simulated states approximating psychedelic, wakeful, conscious, dreaming, sleep, minimally conscious, anaesthetic, non-conscious, and seizure conditions. The index aligns with theoretical expectations and empirical findings, exhibits robust conscious–non-conscious separation synthetically, and tolerates realistic variations in duration, channel count, and noise. The validation of $\Psi$ using both real Sleep-EDF EEG and physiologically informed synthetic EEG demonstrates that the index reliably distinguishes vigilance states and remains stable across biological and model-generated inputs. These findings highlight the theoretical coherence and practical robustness of $\Psi$, positioning it as a promising candidate for low-channel consciousness assessment in both experimental and clinical environments.

By integrating multiscale correlation structure, cross-frequency organisation, and metastable coordination, this framework moves beyond single-dimensional complexity or connectivity measures, offering a richer representation of conscious dynamics. It provides a foundation for future empirical tests using human and animal data and contributes to the toolbox of theory-neutral methods for evaluating and comparing consciousness theories.

\section*{Acknowledgements}
The authors would like to thank the Centre for Visual Computing and Intelligent Systems at the University of Bradford for providing the necessary computing resources to carry out the computational work. 

\section*{Author Contributions}
HU and NH conceived the mathematical model. HU developed the algorithms, ran the computational experiments and documented the results. NH provided input to the analysis the results. Both authors contributed equally to the interpretation of results and manuscript preparation.

\section*{Funding}
No funding was received for this work.

\section*{Availability of data and materials}
Implementation of the consciousness index $\Psi$, including the feature pipeline and synthetic EEG generator, is publicly available on GitHub at: 

\begin{center}
\url{https://github.com/ugail/Index-for-Consciousness-Dynamics}
\end{center}

The code is provided under an open-source license and can be run directly in Google Colab without requiring local installation or specialised hardware. Users can either explore the synthetic validation demonstrations or load their own multi-channel EEG recordings for analysis.

The real EEG recordings used for validation were obtained from the publicly available Sleep-EDF ``Expanded'' dataset, hosted on PhysioNet (\url{https://physionet.org/content/sleep-edfx/}). This dataset is distributed under the Open PhysioNet License and includes overnight polysomnography recordings with manual sleep-stage scoring. Users can freely access and download the data provided they comply with the associated licensing terms.

\section*{Declarations}
\subsection*{Ethics approval and consent to participate}
Not applicable.

\subsection*{Competing interests}
The authors declare no competing interests.

\end{document}